# Resonance vector mode locking

Stanislav A Kolpakov[1], Sergey V Sergeyev[1*], Yuri Loika[1],

Nikita Tarasov[1], Vladimir Kalashnikov[1], Govind P Agrawal[2]

**A mode locked fibre laser as a source of ultra-stable pulse train has revolutionised a wide range of fundamental and applied research areas by offering high peak powers, high repetition rates, femtosecond range pulse widths and a narrow linewidth[1-19]. However, further progress in linewidth narrowing seems to be limited by the complexity of the carrier-envelope phase control[14,15]. Here for the first time we demonstrate experimentally and theoretically a new mechanism of resonance vector self-mode locking where tuning in-cavity birefringence leads to excitation of the longitudinal modes' sidebands accompanied by resonance phase locking sidebands with the adjacent longitudinal modes. An additional resonance with acoustic phonons[18,19] provides the repetition rate tunability and linewidth narrowing down to Hz range that drastically reduces the complexity of the carrier-envelope phase control and so will open the way to advance lasers in the context of applications in metrology, spectroscopy, microwave photonics, astronomy, and telecommunications.**

The mode locked fibre lasers constitute a versatile technology towards producing ultra-stable femtosecond pulse trains (frequency combs having fractional uncertainties of $10^{-19}$ and lower[1-3]) with characteristics required in metrology, high resolution Fourier transform spectroscopy, microwave photonics, remote sensing, astronomy, and telecommunications[1-10]. A wide range of the active and passive mode locking techniques, including phase modulators, additional passive cavities, active fibres with different modal diameters, nonlinear polarisation rotation, and passive saturable absorbers based on carbon nanotubes, graphene, and semiconductor mirrors, have been used to narrow the mode-locked pulse and increase its energy[11-19]. Though these techniques provide pico- or femto-second scale

---

[1] Aston Institute of Photonic Technologies, Aston University, Aston Triangle, Birmingham, B4 7ET, UK
*Corresponding author e-mail address: s.sergeyev@aston.ac.uk
[2] The Institute of Optics, University of Rochester, Goergen 515, Rochester, NY 14627-0186, USA

pulsewidths, it is still the challenge to realize GHz pulse repetition rates with narrow radio-frequency (RF) linewidth in the Hz range[11-19]. Generally the linewidth narrowing techniques are based on the control of the carrier-envelope phase (CEP) $\varphi_{ce}$ and provide zero pulse-to-pulse CEP change ($\Delta\varphi_{ce}=0$) in the envelope phase [1-3] by heterodyning different harmonics i.e. by detecting the slippage rate (beat note) $f_{ce}=f_{rep}\Delta\varphi_{ce}/2\pi$ as a radio-frequency signal (here $f_{rep}$ is the pulse repetition rate). The phase locking of $f_{ce}$ with a reference signal enables CEP stabilisation[1,2]. Such techniques allow the RF linewidth suppression from 1 KHz to 1 mHz but their extraordinary complexity impose limits for applications[1-3].

One way to reduce the complexity of CEP control has been outlined by Grudinin, Gray and co-workers[18-19]. They demonstrated that resonance of a harmonic of the fundamental longitudinal mode with a transverse acoustic wave leads to tunable mode locking with repetition rates between 100 MHz and a few GHz and narrows RF linewidth down to 100 Hz. It has also been found previously that the multimode Risken-Nummedal-Graham-Haken (RNGH) instability[22-25] in an erbium doped fibre laser (EDFL) is the essential mechanism of unstable self-mode-locking. Finding mechanisms beyond the traditional passive or active mode locking techniques for further pulse train stabilisation may result in advancing the frequency comb technology in the context of aforementioned applications.

In this Letter, for the first time we demonstrate theoretically and experimentally new resonance vector mode locking mechanism leading to the pulse train stabilisation. Complemented by the resonance between a harmonic of the longitudinal mode and an acoustic wave excited by this comb through the electrostriction effect, the dynamics finally results in tunability of the repetition rate and linewidth narrowing.

The output power versus pump power, the emission spectrum, and the pulse train are shown in Figure 1b to 1d. When the pump power exceeded 48 mW, stable mode-locked pulses could be observed on the oscilloscope. As shown by Kalashnikov and co-workers[17], observed Lorentzian shape of the optical spectrum in Fig. 1c indicates the presence of chirped pulses. As shown in Figure 1d, the observed pulse train has the fundamental repetition rate of 12.21 MHz. The mean value of linewidth at the fundamental frequency at the 3dB level was only 370 Hz (see INSET 1 of Figure 1d and Table 1). This value is much less than typical values of 10 KHz found for mode locked lasers with a saturable absorber [11-16]. The transient time for stabilisation of this regime varies from a fraction of second to few minutes.

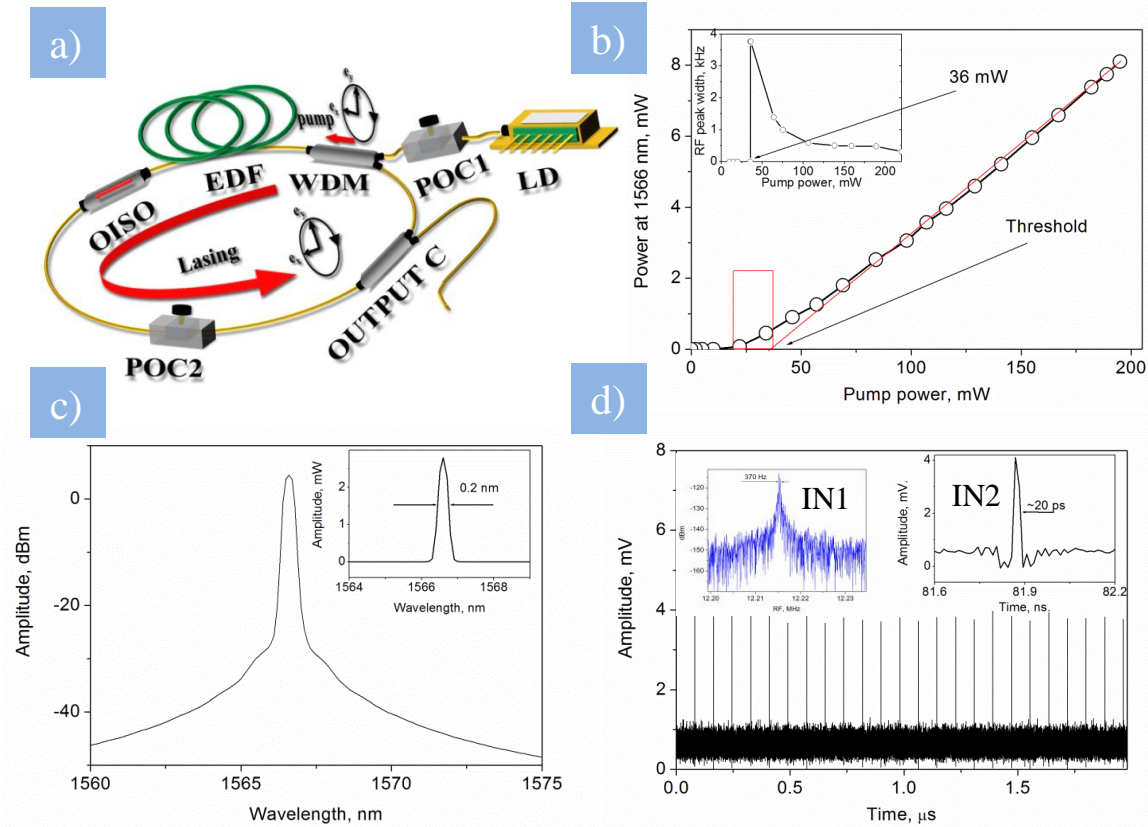

**Figure 1** a, Erbium doped fibre laser. EDF: erbium fibre; LD: l480 nm laser diode for pump; POC1 and POC2: polarization controllers, OISO: optical isolator; WDM: wavelength division multiplexer (WDM), OUTPUT C: 80:20 output coupler. b, Average laser output power versus pump power; INSET: the RF linewidth versus pump power. The rectangle indicates the interval where unstable mode-locking patterns have been observed. c, The optical spectrum, INSET: the same spectra plotted using a linear scale. d, The train of pulses at the fundamental frequency, INSET1: RF spectrum for the fundamental frequency, INSET2: time resolved pulse.

The pulse trace with 20 ps pulsewidth is shown in the INSET 2 of Figure 1d (Details of pulse width measurements are found in section Methods). The most stable patterns observed in our experiments were at the fundamental frequency of 12.21 MHz and its high-order harmonics at frequencies of 293.16 MHz, 464.17 MHz and 549.7 MHz (Table I). The dynamics of the harmonic mode locking at 293.16 MHz is illustrated in Figures 2a-2d. A segment of the RF spectra in the mode locked regime is shown in Figure 2a. The lines "A","B" and "C" correspond to the 23$^{rd}$, 24$^{th}$ and 25$^{th}$ harmonics of the fundamental frequency whereas lines "s" are satellites of the lines "A","B" and "C". To understand the origin Table 1 Frequencies observed in the experiments

| Frequency, MHz. | RF peak width, Hz. | Temporal jitter, ppm[3] | Long term drift |
|---|---|---|---|
| 12,21 | [210, 370, 530] [1,2] | 40 | Yes |
| 97.7 | Unstable | Unstable | - |
| 207.6 | Unstable | Unstable | - |
| 293.16 | [9, 38, 155] [1] | 1.4 | Yes |
| 464.17 | [22, 38, 150] [1] | 0.9 | Yes |
| 549.7 | [1, 13, 97] [1] | 0.5 | Yes |
| 842.5 | Unstable | Unstable | - |
| 903.5 | Unstable | Unstable | - |

[1] *Asymmetric interval of confidence 0. 95 [min, mean, max]*

[2] *At pump power of 220 mW*

[3] *Parts per million relative to the main value of frequency*

of these satellites, we changed birefringence in the laser cavity by turning the knob of the polarization controller POC2 and kept the pump power fixed at 160 mW.

While the angle of the knob was tuned between 18 positions, the satellites of the adjacent lines "A" and "C" were moving closer to the line labelled "B" as shown in Figure 2b. To get insight into the linewidth compression we show temporal traces and RF spectra for the last four steps (labelled with (15), (16), (17) and (18)) in Figures 2c and 2d, respectively. For the position 15 in Figures 2c and 2d the distance between the satellites is slightly less than 3MHz and correspond to the situation when the satellites vanished completely. The RF line corresponding to the fundamental comb frequency appears unchanged and the RF spectrum also seems be unchanged. In the position 16, the distance between satellites was diminished. In this position, the noise demonstrated a periodic pattern, and the RF spectrum became broader and had "three humps". After the knob of POC2 was turned again (17) the oscilloscope traces (Figure 2c, (17)) showed a regularly modulated oscillations at 293.16 MHz and the bias period close to 20 ns (50 MHz). The RF spectrum now exhibited multiple peaks.

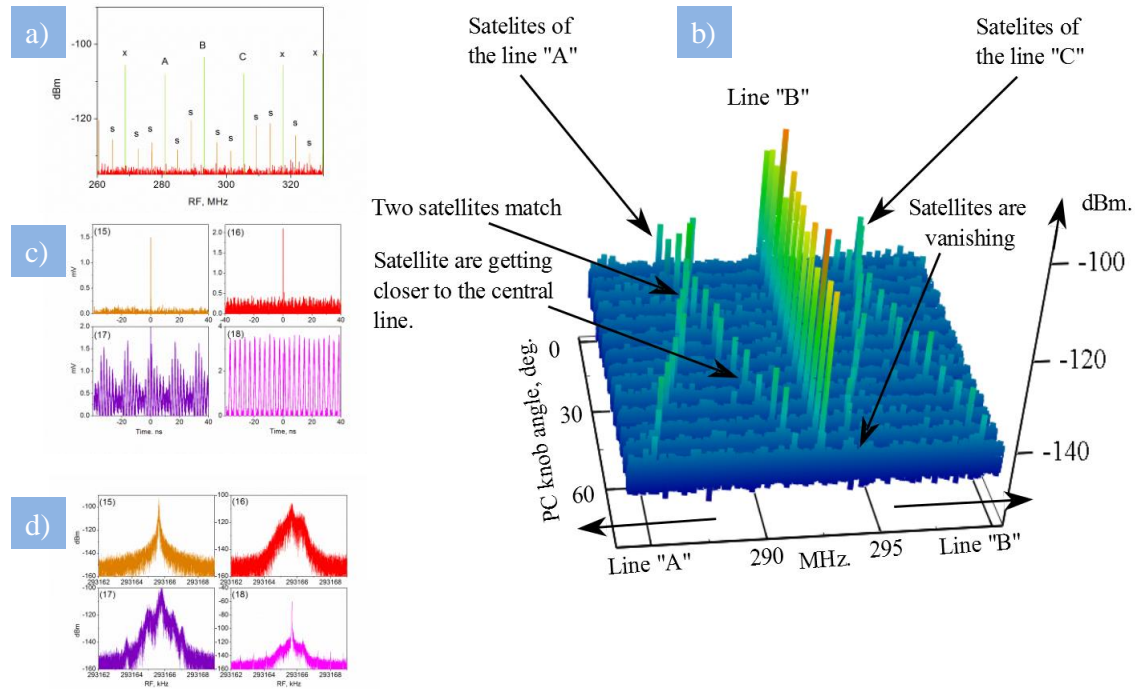

**Figure 2** | a, RF comb showing 23$^{rd}$, 24$^{th}$ and 25$^{th}$ harmonics along with their satellites. b, satellites tuning with the help of in-cavity polarisation controller POC2. c, Emergence of the 293.16 MHz pulse train for the positions 15, 16, 17 and 18 of the POC2. d, Evolution of the RF spectrum of the 293.16 MHz line for the positions 15, 16, 17 and 18 of the POC2.

Finally, after the last rotation of the knob (18) the modulation disappeared and the regular oscillations pattern at the frequency of 293.16 MHz became visible. The RF spectrum now showed a narrow resonance line growing up to 60 dB. In addition to these results, we have observed stable and unstable pulse trains at different frequencies as shown in Table 1 and Supplementary Information. The tuning between different harmonics was performed by adjusting POC2 and the value of the pump power.

To understand the mechanism of stable self-mode locking along with tunability of harmonic mode locking and linewidth narrowing, we have developed a new vector model of EDFLs as described in section Methods. Additionally, the results of linear stability analysis are shown in Fig.3. As shown in section Methods, birefringence tuned by POC2 causes modulation of Stokes parameters along the cavity with an angular frequency equal to the difference between the in-cavity and polarisation hole burning parts of the birefringence strength. In the same way as for the Faraday instability, this spatial modulation is related to the frequency spectrum through the linear dispersion relation defined by the eigenvalue problem and results in the formation of satellites for all longitudinal modes with the splitting proportional to the birefringence strength.

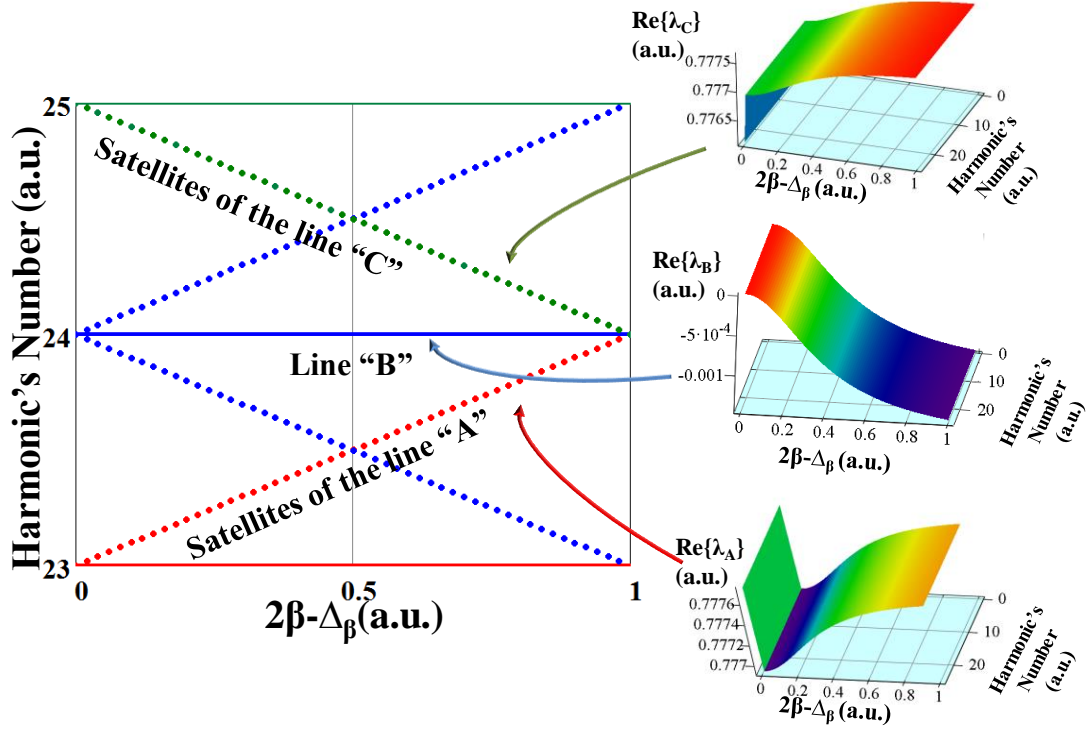

**Figure 3** | Left column, Satellite frequencies for A, B and C lines as a function of the birefringence strength $2\beta$. Frequencies are normalized to the fundamental frequency. Right column, Real part of eigenvalues found from Equations (1). Parameter $\Delta_\beta$ is accounting for birefringence in an active medium induced by polarisation hole burning (details are found in the sections Methods and Supplementary Information).

The same as in Figure 2b, the frequencies of satellites from the adjacent lines "A" and "C" move toward the frequency of line "B" (Figure 3 left column). The mechanism of phase locking is similar to the active mode locking where harmonic modulation of the central frequency $\omega=q$ at the frequency $\pm\Delta\Omega_2$ produces sidebands at the frequencies $q\pm\Delta\Omega_2$ and locks the adjacent modes, which in turn lock their adjacent modes etc. Unlike previous scalar models of RNGH instability, the uniform distribution of the optical field in the cavity is stable with respect to perturbations at the harmonics of the fundamental frequency owing to negative real parts of eigenvalues (Details are found in Supplementary Information). The distinguished feature of the vector model is the presence of the vector branch of complex eigenvalues with the positive real parts that presents evidence of the satellites emergence (Figure 3 in the top and bottom of right column). The interaction of harmonics and satellites beyond the resonance conditions leads to the emergence of the beat tone as illustrated in Figure 2c. Thus, unlike the previous experimentally observed unstable mode locking due to RNGH instability[22-25], we have observed stable harmonic mode locking at 293.16 MHz, 464.17 MHz and 549.7 MHz. Grudinin and Gray [18,19] showed that the stabilisation at these frequencies can be caused by strong coupling between the frequency comb and transverse acoustic wave excited by this comb through electrostriction effect. It seems likely that the synchronisation in the regime of a strong coupling between frequency comb and transverse acoustic wave leads to the phase noise suppression taking the form of the linewidth narrowing [26,27].

In conclusion, we demonstrate experimentally new vector self-mode locking operation of an EDFL. By adjusting the in-cavity polarisation controller and the pump power we were able to switch the stable fundamental mode operation at 12.21 MHz to the harmonic mode locked operation at 293.16 MHz, 464.17 MHz and 549.7 MHz along with linewidth narrowing from hundreds of Hz to a few Hz. To explain stable fundamental mode operation, we developed a new vector model which demonstrates a new kind of the vector multimode RNGH instability. At this point a resonance between the frequency comb and acoustic wave is excited through the electrostriction effect that enforces harmonic mode locking and phase noise suppression taking the form of linewidth narrowing. In the future extension of the vector model given in the section method, we shall develop a vector model of jitter for mode locked fibre lasers to quantify the linewidth narrowing and repetition rate tunability.

**Methods**

**Experimental setup.** The experimental setup is illustrated in Figure 1a. The cavity includes 15.8 m of standard telecommunication fibre (SMF28) and 75 cm of Liekki Er80-8/125 gain fibre (EDF) with the anomalous dispersion of -20 $fs^2$/mm and modal field diameter of 9.5 μm. The length of the whole cavity taking in account the physical length of all components was 17 meters. The pump diode (FOL14xx series with isolator) has the maximum optical power up to 250 mW, which was measured after the polarization controller POC1, optical isolator (not shown in Figure 1a) and wavelength division multiplexer (WDM). A manual polarization controller POC1 and an optical isolator for 1560 nm were placed between the diode output and the WDM.

The isolator between the diode and the laser was used to prevent degradation of the diode power during the device operation and to improve the laser diode stability. The LDR1500E driver was used to drive the laser diode. The output coupler 80:20 was used to direct the light out of the cavity. In the first experiment, the laser was assembled with an optical isolator with 25 dB attenuation. In this configuration it was impossible to obtain periodic pulsations and the output exhibited noisy behaviour. As we found, the back propagated wave interacts with the gain fibre changing the inverse population and preventing stable operation. After installation of an isolator with 51 dB attenuation, the laser was successfully mode locked. The back propagated radiation was controlled through the auxiliary port (not shown in Figure 1a) and the level of the backscattered power was measured to be -49 dB in comparison with the power measured from the port labelled with "OUTPUT C".

**Mode locked operation threshold**. The threshold was found to be close to 36 mW of the pump power using linear extrapolation of the signal vs pump power curve to zero value of the signal power, as illustrated in Figure 1b. When the pump power has the value of 36 mW (inset of Figure 1b) a sharp peak appears at 12.21 MHz in the RF spectrum. As illustrated in the inset, after the peak becomes visible its width decreases from 4 KHz down to 300 Hz.

**Pulse width.** The pulse duration was too large to be measured with an autocorrelator. To estimate the pulse parameters, we used an ultrafast photo-detector XPDV232OR with the bandwidth of 50 GHz. This detector in turn was connected to DSO-X93204A oscilloscope with the bandwidth of 32GHz. The pulse width of 20 ps was obtained using the oscilloscope trace and the $\frac{\sin x}{x}$ interpolation software supplied by Agilent. This algorithm gave us the effective resolution of 781 fs/point. The pulse energy was estimated to be 1.4 nJ. The time bandwidth product (TBWP) was calculated to be 0.5 using the formula $K = \frac{cT\Delta\lambda}{\lambda^2}$, where $T$ is the pulse duration, $\Delta\lambda$ is the width of the optical spectrum and $\lambda$ is central optical wavelength. Since the TBWP for Gaussian pulse is 0.441, we have assumed that pulses have are slightly chirped.

**Vector model.** Evolution of the laser SOPs and population of the first excited level in $Er^{3+}$ doped active medium was modelled using the following equations derived from the vector theory developed by Sergeyev and co-workers in [28,29]:

$$\frac{\partial S_0}{\partial z} + \frac{\partial S_0}{\partial t} = \left(\frac{2\alpha_1 f_1}{1+\Delta^2} - 2\alpha_2\right) S_0 + \frac{2\alpha_1 f_2}{1+\Delta^2} S_1 + \frac{2\alpha_1 f_3}{1+\Delta^2} S_2,$$

$$\frac{\partial S_1}{\partial z} + \frac{\partial S_1}{\partial t} = \gamma S_2 S_3 + \left(\frac{2\alpha_1 f_1}{1+\Delta^2} - 2\alpha_2\right) S_1 + \frac{2\alpha_1 f_2}{1+\Delta^2} S_0 - \frac{2\alpha_1 f_3 \Delta}{1+\Delta^2} S_3,$$

$$\frac{\partial S_2}{\partial z} + \frac{\partial S_2}{\partial t} = -\gamma S_1 S_3 + \frac{2\alpha_1 f_3}{1+\Delta^2} S_0 + \left(\frac{2\alpha_1 f_1}{1+\Delta^2} - 2\alpha_2\right) S_2 + \left(\frac{2\alpha_1 f_2 \Delta}{1+\Delta^2} - 2\beta\right) S_3,$$

$$\frac{\partial S_3}{\partial z} + \frac{\partial S_3}{\partial t} = \frac{2\alpha_1 \Delta f_3}{1+\Delta^2} S_1 - \frac{2\alpha_1 \Delta f_2}{1+\Delta^2} S_2 + 2\beta S_2 + \left(\frac{2\alpha_1 f_1}{1+\Delta^2} - 2\alpha_2\right) S_3,$$

$$\frac{df_1}{dt} = \varepsilon \left[\frac{(\chi-1)I_p}{2} - 1 - \left(1 + \frac{I_p}{2} + d_1 S_0\right) f_1 - \left(d_1 S_1 + \frac{I_p}{2}\frac{(1-\delta^2)}{(1+\delta^2)}\right) f_2 - d_1 S_2 f_3 \right],$$

$$\frac{df_2}{dt} = \varepsilon \left[\frac{(1-\delta^2)}{(1+\delta^2)}\frac{I_p(\chi-1)}{4} - \left(\frac{I_p}{2} + 1 + d_1 S_0\right) f_2 - \left(\frac{(1-\delta^2)}{(1+\delta^2)}\frac{I_p}{2} + d_1 S_1\right)\frac{f_1}{2}\right],$$

$$\frac{df_3}{dt} = -\varepsilon \left[\frac{d_1 S_2 f_1}{2} + \left(\frac{I_p}{2} + 1 + d_1 S_0\right) f_3\right].$$

(1)

Here time and length are normalised to the round trip and cavity length respectively; $S_i$ ($i=0,1,2,3$) are the Stokes parameters ($S_0$ is the output power, pump and lasing powers are normalised to the corresponding saturation powers); $\alpha_1$ is the total absorption of erbium ions at the lasing wavelength, $\alpha_2$ is the total insertion losses in cavity, $\beta$ is the birefringence strength ($2\beta=2\pi L/L_b$, $L_b$ is the beat length); $\delta$ is the ellipticity of the pump wave, $\varepsilon=\tau_R/\tau_{Er}$ is the ratio of the round trip time $\tau_R$ to the lifetime of erbium ions at the first excited level $\tau_{Er}$; $\chi = (\sigma_a^{(L)} + \sigma_e^{(L)})/\sigma_a^{(L)}$, $\sigma_{a(e)}^{(L)}, \sigma_a^{(p)}$ are absorption and emission cross sections at the lasing wavelength and absorption cross section at the pump wavelength; $\Delta$ is the detuning of the lasing wavelength with respect to the maximum of the gain spectrum (normalised to the gain spectral width); $d_1 = \chi/\pi(1+\Delta^2)$. Equations (1) have been derived under approximation that the dipole moments of the absorption and emission transitions for erbium doped silica are located in the plane that is orthogonal to the direction of the light propagation. This results in an angular distribution of the excited ions $n(\theta)$ which can be expanded into a Fourier series as follows:

$$n(\theta) = \frac{n_0}{2} + \sum_{k=1}^{\infty} n_{1k} \cos(k\theta) + \sum_{k=1}^{\infty} n_{2k} \sin(k\theta). \qquad (2)$$

Unlike more general assumption of the 3D orientation distribution of the dipole orientations, Equation (2) allows deriving finite dimension system of the differential equations (1) where only $n_0$, $n_{12}$ and $n_{22}$ components contribute to the vector dynamics. A linear stability analysis of the solution in the case of circularly polarised pump ($\delta=1$), a uniform field ($\partial S_i / \partial z = 0$) and a steady state ($\partial S_i / \partial t = 0$) can be found in Supplementary Information. Using it, we find three branches of eigenstates

$$\begin{aligned}
&(I) \quad \lambda_0 = iq, \\
&(II) \quad \lambda_1 = A_1(q, I_p) + iq, \quad \lambda_2 = A_2(q, I_p) + i(q + \Delta\Omega_1(q, I_p)), \\
&\lambda_3 = A_3(q, I_p) + i(q - \Delta\Omega_1(q, I_p)), \\
&(III) \quad \lambda_4 = A_4(q, I_p, \beta) + iq, \quad \lambda_5 = A_5(q, I_p, \beta) + i(q + \Delta\Omega_2(q, I_p, \beta)), \\
&\lambda_6 = A_6(q, I_p, \beta) + i(q - \Delta\Omega_2(q, I_p, \beta)), \\
&A_1(q, I_p) < 0, \ A_2(q, I_p) < 0, \ A_3(q, I_p) < 0, \ A_4(q, I_p) < 0, \ A_5(q, I_p) > 0, \ A_6(q, I_p) > 0.
\end{aligned} \qquad (3)$$

Here $q$ is the wave number of the longitudinal mode. All details related to numerical calculations of eigenvalues are found in Supplementary Information. For branch III, $q=23, 24,$ and $25$ ($A$, $B$, and $C$ lines and satellites), and $I_p=10$, the particular eigenfrequencies along with real parts of eigenvalues vs birefringence strength are shown in Figure 3. As follows from the structure of Equations (1), birefringence in the laser cavity comprises two parts, viz. birefringence of the passive fibre combined with birefringence induced by the in-cavity polarisation controller ( $2\beta=2\pi L/L_b$ ) and birefringence caused by polarisation hole burning in the active fiber ( $\Delta_\beta=2\alpha_1 f_2 \Delta/(1+\Delta^2)$ ) [28,29]. In the experiments we have observed the threshold pump power of 36 mW. This value is significantly less than the modulation instability threshold. The modulation instability leads to oscillations at frequencies of hundreds GHz [30] which were not observed in the experiment. This means the approximation where the Kerr nonlinearity and second order dispersion are neglected is valid in the context of the qualitative description of the experimentally observed vector self-mode locking. The length of the pulses in the experiments was estimated to be of 20 ps. It is much longer than the time of transverse relaxation of 160 fs. For this reason the dynamics of the medium polarisation also was ignored[22].

ACKNOWLEDGMENTS

We are grateful to Dr Thomas Allsop from the school of Engineering & Applied Science, Aston University, Birmingham for fruitful discussion during the experiments.

This work was supported by Leverhulme Trust (Grant ref: RPG-2014-304), FP7-PEOPLE-2012-IAPP (project GRIFFON, No. 324391) and the West Midlands European Regional Development Fund (ERDF) project.

## Author contributions